\documentstyle[12pt]{article}

\input epsf
\begin{document}
\begin{titlepage}
\today          \hfill
\begin{center}
\hfill    OITS-667 \\

\vskip .05in

{\large \bf Determining the Weak Phase $\gamma$ in the Presence of
Rescattering}
\footnote{This work is supported by DOE Grant DE-FG03-96ER40969.}
\vskip .15in
K. Agashe \footnote{email: agashe@oregon.uoregon.edu} and 
N.G. Deshpande \footnote{email: desh@oregon.uoregon.edu} 
\vskip .1in
{\em
Institute of Theoretical Science \\
5203 University
of Oregon \\
Eugene OR 97403-5203}
\end{center}

\vskip .05in

\begin{abstract}
We suggest a new technique to determine the CKM phase $\gamma$
{\em without} neglecting the (soft) final state rescattering effects. We
use (time integrated) $B$ meson decay rates to $\pi$'s and 
$K$'s.
A set of $5$ $\Delta S = 0$ (or $1$ $\Delta S =0$ and $4$ 
$\Delta S= 1$) 
decay rates is used to compute the
strong phases and magnitudes of the tree level
and penguin contributions
as functions of $\gamma$. These are used to {\em predict}
a $\Delta S = 1$ $(\Delta S = 0)$ $B_{d/s}$ decay rate as a function of 
$\gamma$ (using
$SU(3)$ symmetry). The measurement of this decay rate then gives
$\gamma$. We illustrate this technique using 
different cases.
Most of the decay modes we use are expected to be
accessible at the $B$-factories ($e^+ e^-$ or hadron machines).
\end{abstract}

\end{titlepage}

\newpage
\renewcommand{\thepage}{\arabic{page}}
\setcounter{page}{1}

\section{Introduction}
\label{intro}
%- standard intro. to CP violation in B systems
%The study of CP violation at the $B$-factories will test 
%the Standard Model interpretation of
%CP violation, {\it i.e.,} the single phase
%in the quark mixing, given by the Cabibbo-Kobayashi-Maskawa (CKM) 
%matrix, as the source of CP violation. This is done 
%by (over)determining the parameters
%of the CKM matrix.
%Information on the CKM matrix can be summarized in terms of the
%``unitarity'' triangle which is a representation
%of the unitarity relation between the CKM matrix elements: $V^
%{\star}_{tb} V_{td} +
%V^{\star}_{cb} V_{cd} + V^{\star}_{ub} V_{ud} = 0$, where $V$ is the 
%CKM matrix.
%Thus, 
Determining the angles of the CKM unitarity triangle, denoted by $\alpha$, 
$\beta$ 
and $\gamma$, is one of the important aims of
the $B$-factories. 
Methods have been suggested to determine $\gamma$ ($\equiv$
Arg ($- V^{\star}_{ub} V_{ud} / V^{\star}_{cb} V_{cd})
$ where $V$ is the
CKM matrix) 
using decays of $B_d,B^{+}$
and $B_s$ (and their CP-conjugates) 
to two pseudoscalars belonging to the $SU(3)$ octet 
including
the effects of the Electroweak Penguin (EWP) diagrams.
These methods rely on the flavor $SU(3)$ symmetry.

Many of these methods neglect the effects of (soft) final state
rescattering. In particular, the
decay amplitude for $B^{+} \rightarrow \pi^{+} K^{0}$ is assumed to 
contain only the weak phase $e^{i \pi}$ from 
the penguin diagram with the top quark in the loop.
Tree level operators have the weak phase $e^{i \gamma}$, but since
they have the transition
$\bar{b} \rightarrow
\bar{s} u \bar{u}$, they contribute, in the absence of
rescattering,  
only through annihilation to this decay.
%The reason is as follows. 
%The non-trivial weak
%phase, $e^{i \gamma}$, comes from tree level 
%operators and 
%penguin operators with the up quark in the loop. 
%However, the quark level process for this decay is $\bar{b} \rightarrow
%\bar{s} d \bar{d}$ whereas the 
%tree level operators
%have the transition
%$\bar{b} \rightarrow
%\bar{s} u \bar{u}$. 
%Thus, in the absence of rescattering,
%the tree level
%amplitude 
%contains only the annihilation 
%contribution. 
%%(there are no 
%%spectator quark diagrams due to the tree level operators).
Annihilation  
contributions are argued to 
be small since they are suppressed by $f_B/m_B$ (in the absence of
significant rescattering effects).
%The annihilation diagrams without gluon emissions are also
%helicity suppressed by a factor $m_{u,d,s}/m_B$.
%The up quark penguin diagram has the transition
%$\bar{b} \rightarrow
%\bar{s} d \bar{d}$, but the up quark
%penguin amplitude is suppressed by at least
%the CKM factors
%$\sim |V^{\star}_{ub} V_{us}|/|V^
%{\star}_{tb} V_{ts}| \sim 1/50$ relative to the top quark amplitude. 
%%(again, in
%the short-distance description?). 
%\footnote{Strictly speaking, 
%the penguin
%contribution with the weak phase 
%$e^{i \gamma}$ is $|V^{\star}_{ub} V_{us}|\left( P_u - P_t \right)$
%and the penguin contribution with
%the weak phase $e^{i \pi}$ is 
%$|V^{\star}_{tb} V_{ts}| \left( P_t - P_c \right)$,
%where the $P_q$'s denote
%the penguin amplitudes without the CKM factors. The $P_u$ and $P_c$
%amplitudes can acquire absorptive parts due to on-shell 
%internal up and charm quarks,
%but still there is no compensation of the very large CKM suppression,
%$|V^{\star}_{ub} V_{us}|/|V^
%{\star}_{tb} V_{ts}|$ \cite{bf1}.}
%Thus, if rescattering effects are neglected,
%then the amplitude with the weak phase
%$e^{i\gamma}$ in the decay $B^{+} \rightarrow \pi^{+} K^{0}$ is very small. 
Assuming that
the $B^{+} \rightarrow \pi^{+} K^{0}$ amplitude has no $e^{i \gamma}$ 
weak phase,
references \cite{ghlr1,dh,gr,nr2} have suggested methods to determine 
$\gamma$.

However, rescattering from an intermediate state, for example
$\pi^0 K^+$, 
created 
by a (color-allowed) spectator decay of a $B^+$ 
due to tree level operators,
can generate a significant
amplitude with
the weak phase $e^{i \gamma}$ in the decay $B^+ \rightarrow
\pi^+ K^0$ as well. 
So it would be better to have a method to determine $\gamma$ which
does not use the assumption of no rescattering effects. 
%(and
%hence no $e^{i \gamma}$ weak phase in the decay $B^{+} \rightarrow
%\pi^{+} K^{0}$).
Rescattering might also enhance annihilation
contributions \cite{n} and thus necessitates their inclusion.

Buras and Fleischer \cite{bf2} gave a method to determine
$\gamma$ {\em without} neglecting rescattering using 
$B_{d} \rightarrow \pi^{-}
K^{+}$, $B^{+} \rightarrow \pi^{+} \pi^{0}$ decays and
time {\em dependent} 
measurements of the $B_{d} \rightarrow \pi^{0} K_S$ decay.   
For this method, they also require time dependent analysis 
of, for example, $B_d \rightarrow J/\psi K_S$ to
measure $\beta$.
Gronau and Pirjol \cite{gp} suggested a method using time independent
measurements of {\em all} 
the $B_{d} \rightarrow \pi K$ and $B_s \rightarrow \pi K$
modes. In their method also
rescattering effects are
included. However, it might be diffcult to measure the neutral modes
of $B_s$ decays since that will involve tagging at hadron machines.

In this paper, we suggest a technique to determine $\gamma$ {\em 
including} 
rescattering effects 
(and the EWP operators)
using $B$ meson decays to $\pi$'s and $K$'s. 
We will illustrate
this technique using
four
cases;
see table \ref{cases}.

\renewcommand{\arraystretch}{0.2}
\begin{table}
\begin{center}
\vspace{-1in}
\begin{tabular}
{||c||c|c||}\hline
 & \multicolumn{2}{c||}{ } \\ 
 & \multicolumn{2}{c||}{Modes used} \\ 
 & \multicolumn{2}{c||}{ } \\ \cline{2-3}
 & & \\
Case & $\Delta S =0$ & $\Delta S=1$ \\ 
 & &  \\ \hline
 & & \\
$1$ & $B^+ \rightarrow \pi^+ \pi^0, \;B_d \rightarrow \pi^+ \pi^-, 
\pi^0 \pi^0 
$ & $B_d \rightarrow
\pi^- K^+, \pi^0 K^0$ \\ 
 & &  \\ 
 & $\bar{B}_d \rightarrow
\pi^+ \pi^-, \pi^0 \pi^0$
& $B_s \rightarrow \pi^+ \pi^- \; (\hbox{or} \; \pi^0 \pi^0)
$ \\ 
 & & \\
 & & \\ \hline
 & & \\
$2$ & $B^+ \rightarrow \pi^+ \pi^0, \; B_d \rightarrow \pi^+ \pi^-,
\pi^0 \pi^0$
& $B_s
%\; (\hbox{or} \bar{B}_s) 
\rightarrow K^+ K^-$ \\ 
 & & \\
 & $\bar{B}_d \rightarrow
\pi^+ \pi^-, \pi^0 \pi^0$ & (CP-averaged)\\ 
 & &  \\ 
 & & \\ \hline
 & & \\
$3$ & $B^+ \rightarrow \pi^+ \pi^0, 
\; B_d \rightarrow
\pi^+ \pi^-, \pi^0 \pi^0$
& $
B_d \rightarrow \pi^0 K^0, \pi^- K^+
$ \\  
 & & \\
 & $B_d \rightarrow K^+ K^-$ & 
$\bar{B}_d \rightarrow \pi^0 \bar{K}^0, \pi^+ K^-
$ \\
 & & \\
 & &  \\ \hline
 & & \\
$4$ & $B^+ \rightarrow \pi^+ \pi^0$
 & 
$B_d \rightarrow \pi^0 K^0, \pi^- K^+
$ \\
 & & \\ 
 & $B_s 
%\; (\hbox{or} \bar{B}_s) 
\rightarrow \pi^+ K^- \; (\hbox{or} \; \pi^0 \bar{K}^0)$ (CP-averaged)
 & $\bar{B}_d \rightarrow \pi^0 \bar{K}^0, \pi^+ K^-
$ \\
 & & \\
 & &  \\ \hline
\end{tabular}
\end{center}
\caption{The $6$ (or $8$)
$B$ decay modes used by each of the
$4$ cases to determine
$\gamma$.}
\label{cases}
\end{table} 

We do {\em not}
require any time dependent studies.
The strategy is as follows. In cases \ref{intro} and
\ref{bpipi}, using
$5$ $\Delta S = 0$ decay modes, we determine the strong phases
and magnitudes of the tree level and penguin contributions as functions
of $\gamma$ (assuming flavor $SU(2)$ symmetry). Then, using
flavor $SU(3)$ symmetry, we {\em predict} the rate for {\em one} $\Delta
S = 1$ mode in case \ref{bpipi}.
In case \ref{intro},
two $\Delta S =1$ modes have to be measured to make
a
prediction for a third $\Delta S =1$ mode.
The measurement of the decay for which we have a prediction
(as a function of $\gamma$)
then determines $\gamma$.
A similar idea
can be applied to predict a $\Delta S = 0$ decay mode as a function
of $\gamma$ using measurements of $\Delta S = 1$ (and some $\Delta S =0$)
modes (cases
\ref{bpik} and \ref{conclude}).

%We comment on the accesssibility of these decay modes.
The $B^+ \rightarrow \pi^+ \pi^0$, 
$B_d \rightarrow \pi \pi, \; \pi K$ modes
should be relatively accessible at the $e^+ e^-$ and hadron machines. 
For the $B_d$ decays to a CP eigenstate, 
we require (external)
tagging ({\it i.e.,} the CP-averaged decay rate is not sufficient).
The $B_s$ decay modes were accessible at LEP1 and will be accessible
at hadron machines.
The $B_s \rightarrow \pi \pi$ 
decay mode (case \ref{intro}) might be hard to measure since
it requies tagging whereas in cases \ref{bpipi} and \ref{conclude}, the
$B_s$ modes are either ``self-tagging'' ($\pi^+ K^-$) or a CP-averaged
measurement is sufficient (for $K^- K^+$, $\pi^0 \bar{K}^0$). 
In case
\ref{intro}, we show that if we measure additional $B_d$ modes,
a CP-averaged measurement of the decay rate $B_s \rightarrow \pi \pi$
is sufficient.

Although flavor $SU(3)$ symmetry is used in all the four
cases (as in all the other
methods mentioned above),
in the last section, we discuss how to take into account
$SU(3)$ breaking.

\section{Cases 1 and 2}
\label{bpipi}
We will write the decay amplitudes for the decays $B_i \rightarrow
M \; M$, where $M$ is a pseudoscalar belonging to the
flavor $SU(3)$ octet, in terms of the 5 linearly independent
$SU(3)$ invariant
amplitudes 
%There are six $SU(3)$ invariant
%amplitudes corresponding to the 6 ways of forming
%a flavor $SU(3)$ singlet from $B_i$, the two $M$'s and the effective
%Hamiltonian which transforms as a $\bar{3} \times 3 \times \bar{3}$.
%Of these, the five linearly independent amplitudes 
%are 
denoted by $C_{3}^{T,P}$, $C_{6}^{T,P}$, $C_{15}^{T,P}$, $A_{3}^{T,P}$
and $A_{15}^{T,P}$ (where $T$ and $P$ stand for 
the parts of these amplitudes generated by tree level
and penguin operators, respectively). 
These $SU(3)$ invariant
amplitudes include rescattering effects.
The annihilation amplitudes, $A_{3,15}$, are the ones
in which the quark index $i$ of $B_i$ 
is
contracted directly with the Hamiltonian. 
Neglecting rescattering effects is equivalent to assuming
$C_3^T - C_6^T - C_{15}^T = 0$ and 
$A_{3,15} \sim f_B/m_B$. For example, the tree level part of the
decay amplitude ${\cal A} \left(B^+ \rightarrow \pi^+ K^0
\right)$ contains this
combination of the $C^T$ amplitudes and $A_{15}^T$.

In this notation \cite{d}, 
the amplitudes for $B \rightarrow \pi \pi$
decays 
can be written as
\begin{eqnarray}
- \sqrt{2} {\cal A} (B^+ \rightarrow \pi^+ \pi^0) & = & 8 \;
(\lambda _u ^{(d)} C_{15}^T + \sum_q \lambda _q ^{(d)} C_{15,q}^P ) 
\nonumber \\
 & = & -3 \; I_2, 
\label{b+}
\end{eqnarray}
\begin{eqnarray}
\sqrt{2} {\cal A} (B_d \rightarrow \pi^0 \pi^0) 
& = & \lambda _u ^{(d)} (C_3^T + C_6^T - 
5 C_{15}^T ) + \sum_q \lambda _q ^{(d)} (C_{3,q}^P + C_{6,q}^P - 
5 C_{15,q}^P ) \nonumber \\
 & & + \lambda _u ^{(d)} \left( 2 A_{3}^T + A_{15}^T \right) + \sum_q 
\lambda _q ^{(d)} \left( 
2 A_{3,q}^P + A_{15,q}^P \right) \nonumber \\
 & = & - I_0 + 2 I_2, 
\label{b0}
\end{eqnarray}
\begin{eqnarray} 
{\cal A} (B_d \rightarrow \pi^+ \pi^-) & = & -\lambda _u ^{(d)} 
(C_3^T + C_6^T +
3 C_{15}^T ) - \sum_q \lambda _q ^{(d)} (C_{3,q}^P + C_{6,q}^P +
3 C_{15,q}^P ) \nonumber \\
 & & - \lambda _u ^{(d)} \left( 2 A_{3}^T + A_{15}^T \right) - \sum_q 
\lambda _q ^{(d)} \left(
2 A_{3,q}^P + A_{15,q}^P \right) \nonumber \\
 & = & I_0 + I_2.
\label{b+-}
\end{eqnarray}
Here, $\lambda _q ^{(q^{\prime})} = V^{\star}_{q b} V_{q q^{\prime}}$
($q = u,c,t$ and $q^{\prime} = d,s$) and $C_q^P$, $A^P_q$ denote the penguin 
amplitudes due to $q$ running in the loop. \footnote{Tree level
operators with the flavor structure
$c \bar{c} \; \bar{b} d$ can also
contribute through rescattering from charm intermediate states.
This rescattering generates a charm-quark penguin topology with
the amplitude being proportional to $\lambda _c^{(d)}$ and so can be included
as part of $C_{i,c}^P$.}
$I_2$ and $I_0$ are the amplitudes for $B \rightarrow
\pi \pi (I=2)$
and $(I=0)$ respectively or in other words the $\Delta I = 3/2$ and
$\Delta I = 1/2$ amplitudes.

Using the unitarity of the CKM matrix, {\it i.e.,} $\lambda _t ^{(d)}
= - \lambda _u ^{(d)} - \lambda _c ^{(d)}$, we get
\begin{equation}
\lambda _u ^{(d)} C^T_i + \sum_q \lambda _q ^{(d)} C_{i,q}^P =
\lambda _u ^{(d)} \tilde{C}^T_i - \lambda _c ^{(d)} C_i^P
\label{notation}
\end{equation}
where $\tilde{C}^T _{i} =C_{i}^T - C_{i,t}^P + C_{i,u}^P$ and
$C_{i}^P = C_{i,t}^P - C_{i,c}^P$. 
A similar notation is used for $\tilde{A}_i^T$ and $A_i^P$.  

In the $B^+ \rightarrow \pi^+ \pi^0$ decay, which contains only
the $\Delta I = 3/2$ amplitude, there is no contribution
to $C_{15,q}^P$
from the strong penguin diagrams since these diagrams are $\Delta I = 1/2$.
Neubert and Rosner \cite{nr1} showed that $C_{15,q}^P = C_{15}^T \; 3/2 \;
\kappa _q$,
where $\kappa _q = \; (c_{9,q} + c_{10,q}) / (c_1 + c_2)$ is the ratio
of Wilson coefficients (WC's) of the 
EWP operators (with quark $q$ running in the loop) and the tree 
level operators in the effective Hamiltonian. 
We expect $c_{(9,10),t} \gg
c_{(9,10),(u,c)}$ since the top quark EWP diagram with $Z$ exchange is
enhanced by $m^2_t/m_Z^2$ and so henceforth
we neglect
%$c_{(9,10),(u,c)}$
$\kappa _{u,c}$
and denote $\kappa _t$ by
$\kappa$.
So, we get
\begin{equation}
\tilde{C}_{15}^T \approx C^T_{15} \; \left( 1 - \frac{3}{2} \kappa \right),
\label{nrresult0}
\end{equation}
\begin{equation}
C_{15}^P \approx C_{15}^T \; \frac{3}{2} \kappa
\label{nrresult}
\end{equation}
Using Eqns.(\ref{notation}), (\ref{nrresult0}) and (\ref{nrresult}), 
we can
rewrite $ - 3 \; I_2 = 8 \;
(\lambda _u ^{(d)} C_{15}^T + \sum_q \lambda _q ^{(d)} C_{15,q}^P )$ 
\\ 
$= 8 \;
\left(\lambda _u ^{(d)} \tilde{C} _{15} ^T - \lambda _c ^{(d)}
C_{15}^P \right)$ as
\begin{eqnarray}
- 3\; I_2 & 
\approx & 8 \; \tilde{C} _{15} ^T \left( \lambda _u ^{(d)} - \frac{
\frac{3}{2} \kappa}{1 - \frac{3}{2} \kappa} \lambda _c ^{(d)} \right).
\label{a2}
\end{eqnarray}
Since $3/2 \; \kappa \sim 2\%$ and $|\lambda _u ^{(d)}| 
\sim |\lambda _c ^{(d)}|$, we neglect the second term
({\it i.e.}, the EWP contribution) in the right hand side of
Eqn.(\ref{a2})
for now and we assume $|\tilde{C}_{15}^T| \approx
|C_{15}^T|$. We will show later how to include it.
Then, using the Wolfenstein parametrization in which
$\lambda _u ^{(d)} = |\lambda _u ^{(d)}| e^{i \gamma}$
and $\lambda _c ^{(d)} = |\lambda _c ^{(d)}|$ (and similarly
for $d$ replaced by $s$), we get
\begin{eqnarray}
-3 \; I_2 \approx |\lambda _u ^{(d)}| e^{i \gamma} \; 8 \; 
|\tilde{C}_{15}^T|
\approx |\lambda _u ^{(d)}| e^{i \gamma} \; 8 \; |C_{15}^T|
\label{approxa2}
\end{eqnarray}
so that $|C_{15}^T|$ can be obtained directly
from the $B^+ \rightarrow \pi^+ \pi^0$ 
decay rate. We have chosen a phase convention such that
the strong phase of $C_{15}^T$ is zero.
 
From Eqns.(\ref{b0}) and (\ref{b+-}), we get
\begin{eqnarray}
I_0 & = & - \lambda _u ^{(d)} \left(\tilde{C}_3^T + \tilde{C}_6^T + 
\frac{1}{3} \tilde{C}_{15}^T \right) + \lambda_c^{(d)} (C_3^P + C_6^P + 
\frac{1}{3}
C_{15}^P) \nonumber \\
 & & - \lambda _u ^{(d)} \left( 2 \; \tilde{A}_{3}^T + \tilde{A}_{15}^T 
\right) +
\lambda _c ^{(d)} \left(
2 A_{3}^P + A_{15}^P \right) \nonumber \\
 & \equiv & e^{i \phi_{\tilde{T}}} \; |\lambda _u ^{(d)}| \; e^{i \gamma} 
\; \tilde{T}
- |\lambda _c ^{(d)}| \; e^{i \phi_P} \; P.
\label{a0}
\end{eqnarray}

The five quantities: $|C_{15}^T|$, 
$\tilde{T}$, $P$, $\phi_{\tilde{T}}$ and $\phi_P$ 
(where the $\phi$'s are the CP conserving strong phases)
can thus be determined
as functions of $\gamma$
from the measurements of the five rates: $B^+ \rightarrow \pi^+ \pi^0$, 
$B_d \rightarrow \pi^+ \pi^-$, $B_d \rightarrow \pi^0 \pi^0$ and 
the CP-conjugates
of the $B_d$ decays. 

\begin{figure}
\vspace{-1in}
\centerline{\epsfxsize=0.8\textwidth \epsfbox{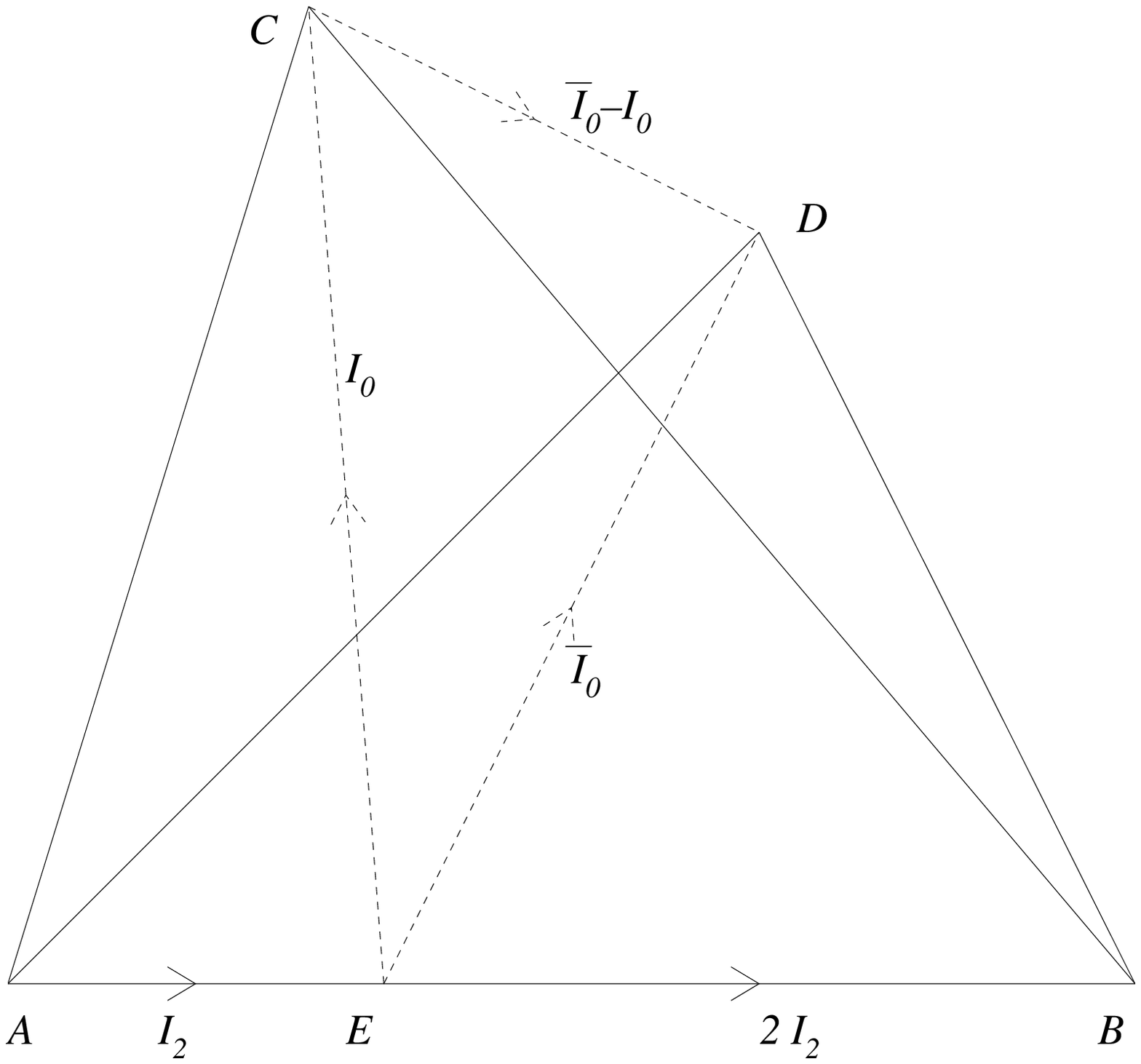}}
\vspace{-0.8in}
\caption{
The triangles formed by the $B \rightarrow \pi \pi$
amplitudes: $AB = |
\protect\sqrt{2}
{\cal A} \left( B^+ \rightarrow \pi^+ \pi^0 \right)|$,
$AC = |{\cal A} \left( B_d \rightarrow \pi^+ \pi^-  \right)|$, $BC = |
\protect\sqrt{2} {\cal A}
\left( B_d \rightarrow
\pi^0 \pi^0 \right) |$, $AD = | {\cal A} \left( \bar{B}_d
\rightarrow \pi^+ \pi^- \right)| $ and $BD = |
\protect\sqrt{2} {\cal A} \left(
\bar{B}_d \rightarrow
\pi^0 \pi^0 \right)|$. In the
%coordinate
%system
phase convention
where the strong phase of $C_{15}^T$ is zero, the angle
between $I_2$ and the real axis (not shown)
is $\pi + \gamma$ (see Eqn.(\protect\ref{approxa2})).}
\protect\label{figbpipi}
\end{figure}

Explicitly, rotating the CP-conjugate amplitudes by $e^{i 2 \gamma}$
(and denoting them by ``bars''),  
we get the triangle formed by $B^+ \rightarrow \pi^+ \pi^0$,
$B_d \rightarrow \pi^+ \pi^-$ and $B_d \rightarrow \pi^0 \pi^0$ 
(Eqns.(\ref{b+}),
(\ref{b0}) and (\ref{b+-})):
\begin{equation}
-\sqrt{2}{\cal A}(B^+ \rightarrow \pi^+ \pi^0) + \sqrt{2}{\cal A}(B_d 
\rightarrow
\pi^0 \pi^0) + {\cal A}(B_d \rightarrow
\pi^+ \pi^-) = 0
\end{equation}
and the one formed by the CP-conjugate decays. These are 
shown in Fig.\ref{figbpipi} (from 
Eqn.(\ref{approxa2}), $I_2 = \bar{I}_2$).

From Eqn.(\ref{a0}), we get
\begin{equation}
\bar{I}_0 - I_0 = - |\lambda _c ^{(d)}| e^{i \phi_P} P \left(e^{i 2 
\gamma} -1
\right).
\label{idiff1}
\end{equation}
Thus, the length of $\bar{I}_0 - I_0$ (obtained from
Fig.\ref{figbpipi})
gives $P$ as a function of $\gamma$.
The angle between $\bar{I}_0 - I_0$ and $I_2$ is 
$\phi _P + \pi /2$ (see Eqns.(\ref{approxa2}) and
(\ref{idiff1})) so that the orientation of $\bar{I}_0 - I_0$
(obtained from Fig.\ref{figbpipi}) gives $\phi _P$ independent of
$\gamma$. \footnote{
This was also discussed in \cite{nq}.}

Similarly,
\begin{equation}
\bar{I}_0 e^{- i 2 \gamma} - I_0
= |\lambda _u ^{(d)}| e^{i \phi_{\tilde{T}}} \tilde{T} \left(e^{ - i
\gamma}
- e^{i \gamma}
\right)
\label{idiff2}
\end{equation}
gives $\phi_{\tilde{T}}$ and $\tilde{T}$ (or knowing
$I_0, \;P$ and $\phi_P$ gives $\phi_{\tilde{T}}$ and $\tilde{T}$
using Eqn.(\ref{a0})).

%A priori, 
There is a discrete ambiguity in this procedure
since in Fig.\ref{figbpipi}, the vertices $C$ and $D$
could be on opposite sides of $I_{2}$. 
This has been discussed in the literature \cite{nq} \cite{gl}.
%That this possibility is
%unlikely can be seen as follows. We expect $|I_0 - \bar{I}_0|
%\sim |\lambda _c ^{(d)}| P$ to be smaller than the tree level amplitude,
%say, $|C^T_{15}| \sim |I_{2}|$. Also, we expect the tree level 
%contribution to
%dominate in the $B_d \rightarrow \pi^+ \pi^-$ decay so that
%the angle between the sides $AC \; \left( {\cal A} \left( 
%B_d \rightarrow
%\pi^- \pi^+ \right) \right)$ and $AD \; \left(
%e^{i 2 \gamma} \;  {\cal A} \left(
%\bar{B}_d \rightarrow
%\pi^- \pi^+ \right) \right)$ is small. Both
%these expectations are consistent with the orientations of the
%two triangles $ABC$ and $ABD$ shown in Fig.\ref{figbpipi}
%and {\em in}consistent if one of the triangles is flipped
%about the side $AB$.

All the analysis up to now
actually relies only on flavor
$SU(2)$ symmetry.

\subsection{Case 1}
The $B_d \rightarrow K \pi$ and $B_s \rightarrow \pi \pi$
amplitudes are given by \cite{d}
\begin{eqnarray}
- {\cal A} (B_d \rightarrow K^+ \pi^-) & = & 
\lambda _u ^{(s)} (C_3^T + C_6^T+ 
3 C_{15}^T ) + \sum _q \lambda _q ^{(s)} \left(C_{3,q}^P + C_{6,q}^P+ 
3 C_{15,q}^P \right) \nonumber \\
 & & - \lambda _u ^{(s)} A_{15}^T - \sum _q \lambda _q^{(s)} A_{15,q}^P,
\label{bk+pi-}
\end{eqnarray}
\begin{eqnarray}
\sqrt{2} \;  {\cal A} (B_d \rightarrow
\pi^0  K^0) & = &
\lambda _u ^{(s)} (C_3^T + C_6^T -
5 C_{15}^T ) + \sum _q \lambda _q ^{(s)} \left(C_{3,q}^P + C_{6,q}^P -
5 C_{15,q}^P \right)
\nonumber \\
 & & - \lambda _u ^{(s)} A_{15}^T - \sum _q \lambda _q^{(s)} A_{15,q}^P,
\label{bk0pi0}
\end{eqnarray}
\begin{eqnarray}
2 \; {\cal A} \left( B_s \rightarrow \pi^+ \pi^- \right) & = & {\cal A} 
\left( B_s \rightarrow \pi^0 \pi^0 \right) \nonumber \\
 & = & - 
\lambda _u ^{(s)} \left( 4 \; A_{3}^T + 4 \; A_{15}^T \right) 
\nonumber \\
 & & - \sum_q
\lambda _q ^{(s)} \left(
4 \;  A_{3,q}^P + 4 \; A_{15,q}^P \right).
\label{bspipi}
\end{eqnarray}
From Eqns.(\ref{bk+pi-}) and (\ref{bspipi}) 
and using the notation
of
Eqn.(\ref{notation}) and rearranging we get
\begin{eqnarray}
{\cal A} (B_d \rightarrow K^+ \pi^-) + {\cal A} \left( B_s \rightarrow
\pi^+ \pi^- \right)
 & = & - \lambda _u ^{(s)} \left(\tilde{C}_3^T + \tilde{C}_6^T +
\frac{1}{3} \tilde{C}_{15}^T \right)
\nonumber \\
 & & + \lambda_c^{(s)}
(C_3^P + C_6^P
+ \frac{1}{3}
C_{15}^P)  \nonumber \\
 & & - \lambda _u ^{(s)} \left( 2 \; \tilde{A}_{3}^T + \tilde{A}_{15}^T
\right) +
\lambda _c ^{(s)} \left(
2 A_{3}^P + A_{15}^P \right) \nonumber \\
 & & - \frac{8}{3} \left( \lambda _u ^{(s)} \tilde{C}_{15}^T
- \lambda_c^{(s)} C_{15}^P \right).
\label{bkpibspipi}
\end{eqnarray}
Using Eqns.(\ref{nrresult}) and (\ref{a0}) 
%the notation 
%of 
%Eqn.(\ref{notation})
%$\tilde{C}^T_i$, $C^P_i$, $\tilde{A}^T_i$, $A^P_i$ 
%defined earlier
and assuming
$C_{15}^T \approx \tilde{C}^T_{15}$, we get 
\begin{eqnarray}
{\cal A} (B_d \rightarrow K^+ \pi^-) + {\cal A} \left( B_s \rightarrow
\pi^+ \pi^- \right) & = & 
e^{i \phi_{\tilde{T}}} |\lambda _u ^{(s)}| e^{i \gamma} \tilde{T}
 -  |\lambda _c ^{(s)}| e^{i \phi_P} P
\nonumber \\
 & & -  \frac{8}{3} C_{15}^T \left( \lambda _u ^{(s)} - \frac{3}{2} \;
\kappa \;
\lambda _c ^{(s)} \right),
\label{bkpibspipi1}
\end{eqnarray}
{\it i.e.,} from Eqns.(\ref{b+-}) and (\ref{bkpibspipi}), we
see that 
the combination of the amplitudes
${\cal A} (B_d \rightarrow K^+ \pi^-) + {\cal A} 
\left( B_s \rightarrow
\pi^+ \pi^- \right)$ can be obtained from the 
amplitude for $B_d \rightarrow  \pi^+ \pi^-$ by
scaling the tree level contribution in the latter by
$|\lambda _u ^{(s)}|/|\lambda _u ^{(d)}|$ and the penguin contribution by 
$|\lambda _c ^{(s)}|/|\lambda _c ^{(d)}|$. 
%\footnote{
%We need to assume flavor $SU(3)$ symmetry here, 
%{\it i.e.,}
%we assume that the amplitudes 
%%$C_{15}^{T,P}$, $\tilde{T}
%%e^{\phi_{\tilde{T}}}$ and $P e^{\phi _P}$
%$C_i^{T,P}$
%are the same for $\Delta S=0$ and $\Delta S =1$ decays. 
%} 
In particular
the EWP contribution ($\propto \lambda _c ^{(s)}$)
is important in the last line of Eqn.(\ref{bkpibspipi1}) 
since, due to the CKM factors, 
it is comparable to the tree level contribution
(unlike in the $B_d \rightarrow  \pi^+ \pi^-$
decay; see Eqn.(\ref{a2})).

Similarly, from Eqns.(\ref{b0}), 
(\ref{bk0pi0}) and (\ref{bspipi}), we see that
the combination of the amplitudes $\sqrt{2} \; 
{\cal A} (B_d \rightarrow K^0 \pi^0) - {\cal A}
\left( B_s \rightarrow
\pi^- \pi^+ \right)$ can be obtained from the
amplitude $\sqrt{2} 
\; {\cal A} \left( 
B_d \rightarrow  \pi^0 \pi^0 \right)$ by scaling the latter by
CKM factors. This gives  
\begin{eqnarray}
\sqrt{2} \; 
{\cal A} (B_d \rightarrow K^0 \pi^0) - {\cal A} \left( B_s \rightarrow
\pi^+ \pi^- \right) & = &
- e^{i \phi_{\tilde{T}}} |\lambda _u ^{(s)}| e^{i \gamma} \tilde{T}
+  |\lambda _c ^{(s)}| e^{i \phi_P} P
\nonumber \\
 & & - \frac{16}{3} C_{15}^T \left( \lambda _u ^{(s)} - \frac{3}{2}
\;  \kappa \;
\lambda _c ^{(s)} \right).
\label{bkpibspipi2}
\end{eqnarray}
We emphasize again that in obtaining the last lines of 
Eqns.(\ref{bkpibspipi1}) and 
(\ref{bkpibspipi2}), it is 
crucial that we use the Neubert-Rosner result (Eqn.(\ref{nrresult})),
{\it i.e.,} that $\kappa$ is calculable.
If
${\cal A} \left( B_s \rightarrow \pi^+ \pi^-
\right) \equiv a^{\prime} 
\; e^{i \phi ^{\prime} _a}$, we can determine
$a^{\prime}$ from the decay rate $B_s \rightarrow
\pi^+ \pi^-$.
Then, measuring the decay rate $B_d \rightarrow \pi^- K^+$
gives $\phi ^{\prime} _a$ as a 
function of $\gamma$ (using Eqn.(\ref{bkpibspipi1})) ($\phi _P$,
$\phi _{\tilde{T}}$, $P$ and $\tilde{T}$ are already known
as functions of $\gamma$).
Knowing $a^{\prime}$ and $\phi ^{\prime} _a$,
we have a prediction 
for the decay rate
$B_d \rightarrow K^0 \pi^0$ (Eqn.(\ref{bkpibspipi2}))
and then $\gamma$ can be determined by measuring this decay. 
\footnote{In this part of the technique, we introduce two additional
discrete ambiguities -- one in determining
$\phi ^{\prime} _a$ and another in the final determination of 
$\gamma$ from the $B_d \rightarrow K^0 \pi^0$ rate.} Thus, we
can determine $\gamma$, including rescattering effects,
by measuring the  $8$ decay modes: $B^+ \rightarrow \pi^+ \pi^0$,
$B_d$ and $\bar{B}_d \rightarrow \pi^+ \pi^-, \pi^0 \pi^0$, 
$B_d \rightarrow \pi^- K^+$, $B_d \rightarrow K^0 \pi^0$ and 
$B_s \rightarrow \pi \pi$ (any one) (or
CP-conjugates of the last three modes). 

From Eqn.(\ref{bspipi}), we see that the decay mode $B_s \rightarrow
\pi \pi$ has only an annihilation contribution.
If 
(either from experimental
measurement of the $B_s \rightarrow \pi \pi$ rate (or a limit on the rate) 
or a theoretical estimate including
rescattering) 
the 
annihilation amplitude $\sim A_3 + A_{15}$ does turn out to be small (smaller
than, 
say, the experimental error in the measurement of the
(magnitude) of ${\cal A} \left( B_d \rightarrow \pi K \right)$) 
then, a decay rate $B_d 
\rightarrow \pi K$ can be predicted
(as a function of $\gamma$)
by simply scaling the corresponding
$B_d \rightarrow \pi \pi$ amplitude by CKM factors. 
Thus, in this case,
$6$ decay modes: $B^+ \rightarrow \pi^+ \pi^0$,
$B_d$ (and $\bar{B}_d$) $\rightarrow \pi^+ \pi^-,
\pi^0 \pi^0$ and any {\em one}
$B_d \rightarrow \pi K$ are sufficient to determine $\gamma$.

As mentioned in the introduction, we require tagging to measure the
$B_s \rightarrow
\pi \pi$ decay mode ({\it i.e.}, the CP-averaged rate is not
sufficient in the above method). If this tagged decay rate is hard 
to measure
whereas the CP-averaged rate can be measured, we can proceed as follows.
Writing 
${\cal A} \left( \bar{B}_s \rightarrow \pi^+ \pi^- \right) 
\equiv \bar{a}^{\prime} e^{\bar{\phi}_a^{\prime}}$, the CP-averaged rate
is $1/2 \left( a^{\prime \; 2} + \bar{a}^{\prime \; 2} \right)$. Measuring
the $B_d \rightarrow \pi K$ decay rates, from 
Eqns.(\ref{bkpibspipi1})
and (\ref{bkpibspipi2}), we can determine $a^{\prime}$ 
(and $\phi_a^{\prime}$) as a function
of $\gamma$. Similarly, the 
%CP-conjugate 
$\bar{B}_d
\rightarrow \pi K$ decay rates will give
$\bar{a}^{\prime}$. Thus, the CP-averaged 
$B_s \rightarrow
\pi \pi$ rate can be predicted as a 
function of $\gamma$ and its measurement gives $\gamma$.

\subsection{Case 2}
The expression for the $B_s \rightarrow K^+ K^-$ decay
amplitude in terms of the
$SU(3)$ invariant amplitudes 
is identical to that for $B_d \rightarrow \pi^- \pi^+$, {\em
including} annihilation contributions (unlike the decay mode
$B_d \rightarrow \pi^- K^+$),
modulo CKM factors, {\it i.e.,}
up to $\lambda^{(d)}_q \rightarrow \lambda^{(s)}_q$ \cite{d}. 
Thus, we also 
have a 
prediction (as a function of $\gamma$)
for this decay rate and also the rate
for its CP-conjugate process including {\em all} rescattering effects. 
So,
the measurement of this CP-averaged decay rate
can be used to determine $\gamma$.

\section{Cases 3 and 4}
\label{bpik}
Using the same technique as in section \ref{bpipi},
we can {\em predict} the rate for a $B_d \rightarrow \pi \pi$ 
(or a $B_s \rightarrow \pi K$) decay
as a function of $\gamma$ {\em given} the decay rates for $B_d \rightarrow
\pi K$. 
The difference is that 
in the triangle construction to determine the tree level and penguin
contributions to the $B_d \rightarrow \pi K$ amplitudes,
we have to take into account
the EWP contributions to the $B_d \rightarrow
\pi K$ amplitudes (the EWP amplitudes
were neglected in the $B \rightarrow \pi \pi$
triangles in section \ref{bpipi}; see Eqns.(\ref{a2})
and (\ref{approxa2})).
\footnote{
Of course, in section \ref{bpipi}, to make a {\em prediction} for
$\Delta S =1$ decay modes, we did have to include the EWP contribution to the 
$\Delta S =1$ decay amplitudes.}
So, we discuss the application of the technique again.

The decay amplitudes for $B_d \rightarrow
\pi  K$ (Eqns.(\ref{bk+pi-}) and (\ref{bk0pi0})) can be written as 
\begin{equation}
\sqrt{2} \; {\cal A} (B_d \rightarrow
\pi^0  K^0)   =  - I_{1/2} + 2 I_{3/2}, 
\end{equation}
\begin{equation}
{\cal A} (B_d \rightarrow
\pi^-  K^+) = I_{1/2} + I_{3/2},
\label{bk+pi-2}
\end{equation}
where $I_{1/2}$ and $I_{3/2}$ are the amplitudes for 
$B_d$ decay to $\pi \; K \; (I=1/2)$ and $(I = 3/2)$ respectively.
Then,
\begin{eqnarray}
3 I_{3/2} & = & \sqrt{2} {\cal A}(B_d \rightarrow
\pi^0  K^0) + {\cal A}(B_d \rightarrow
\pi^-  K^+) \nonumber \\
 & = & - \lambda _u ^{(s)} 8 \tilde{C} _{15}^T + 8 \lambda _c ^{(s)}
C_{15}^P \nonumber \\
 & = & -8 C_{15}^T \left(\lambda _u ^{(s)}
 - \frac{3}{2} \kappa 
\lambda _c ^{(s)} \right) \nonumber \\
 & = & -8 \; |C_{15}^T| |\lambda _u ^{(s)}| \left(e^{i \gamma} + 
\delta _{EW} \right),
\label{A3/2}
\end{eqnarray}
using
%\footnote{
%We need to use $SU(3)$ symmetry here, {\it i.e.,} we assume
%that the $C_{15}^{T,P}$ amplitudes are the same for the
%$B_d \rightarrow \pi K$ and $B^+ \rightarrow \pi^+ \pi^0$ decays.}
Eqn.(\ref{nrresult})
%$C_{15}^P = C_{15}^T \; 3/2 \; \kappa$
and $| \tilde{C} _{15}^T 
|\approx
| C_{15}^T |$ ({\it i.e.,} neglecting the EWP contribution in the
$B^+ \rightarrow \pi^+ \pi^0$ decay). 
$\delta _{EW}$ is given by $-|\lambda _c ^{(s)}| 
/ |\lambda _u ^{(s)}| \; 3/2 \; \kappa 
\sim O(1) $, {\it i.e.,} as mentioned 
earlier,
the EWP contribution {\em is} important for $B_d \rightarrow \pi K$
decays.
$|C_{15}^T|$ can be obtained from the $B^+ \rightarrow \pi^+ \pi^0$ decay
rate as before.

$I_{1/2}$ is 
given by (in analogy to $I_0$ of section \ref{bpipi})
\begin{eqnarray}
I_{1/2} & = & -\lambda _u ^{(s)} \left(\tilde{C}_3^T + \tilde{C}_6^T +
\frac{1}{3} \tilde{C}_{15}^T \right) + \lambda_c^{(s)} (C_3^P + C_6^P + 
\frac{1}{
3}
C_{15}^P) 
\nonumber \\
 & & + \lambda _u ^{(s)} \tilde{A}_{15}^T -  \lambda _c^{(s)} A_{15}^P
\nonumber \\
 & \equiv & e^{i \phi^{\prime}_{\tilde{T}}} |\lambda _u ^{(s)}| 
e^{i \gamma} \tilde{T}^{\prime}
- |\lambda _c ^{(s)}| e^{i \phi^{\prime}_P} P^{\prime}.
\label{a1/2}
\end{eqnarray}

As in section \ref{bpipi}, the four quantities: 
$\tilde{T}^{\prime}$, $P^{\prime}$, $\phi^{\prime}
_{\tilde{T}}$ and $\phi^{\prime}_P$ can thus be determined
as functions of $\gamma$
from the measurements of the four decay rates:
$B_d \rightarrow \pi^- K^+$, $B_d \rightarrow \pi^0 K^0$ and their
CP-conjugates. \footnote{A simlilar analysis can be done with the
$B^+ \rightarrow \pi K$ decay amplitudes which can be written in terms
of $I_{3/2}$ (the same as for $B_d \rightarrow \pi K$ decays)
and $I^{\prime}_{1/2}$ which is a different combination of
the $SU(3)$ invariant amplitudes than $I_{1/2}$. Thus, the $B^+ \rightarrow
\pi K$ decay rates are not useful as far as predicting the $B_d 
\rightarrow \pi \pi$
(case \ref{bpik})
(or $B_s \rightarrow \pi K$ in case \ref{conclude}) decay rates is concerned.}

Due to the EWP contribution (see Eqn.(\ref{A3/2})),
the triangle construction is a bit different in this case as shown below.

As before, we multiply the CP-conjugate amplitudes by
$e^{i 2 \gamma}$ to get the ``barred'' amplitudes.
In this case 
(unlike the case for
$I_2$ in section \ref{bpipi}) 
there is an angle between $I_{3/2}$ and $\bar{I} _{3/2}$
denoted by
$2 \tilde{\gamma}$
and their magnitudes 
are functions of $\gamma$ (see Eqn.(\ref{A3/2})):
\begin{equation}
|I_{3/2}| = |\bar{I} _{3/2}| = \frac{8}{3} |C_{15}^T| |\lambda _u ^{(s)}|
\sqrt{ \left( 1 + \delta _{EW} ^2 + 2 \delta _{EW} \cos \gamma \right) }
\label{maga3/2},
\end{equation}
\begin{equation}
\tan \tilde{\gamma} = \frac{\delta _{EW} \sin \gamma}{1 + \delta _{EW} 
\cos
\gamma}.
\label{anga3/2}
\end{equation}

\begin{figure}
\vspace{-1.7in}
\centerline{\epsfxsize=0.8\textwidth \epsfbox{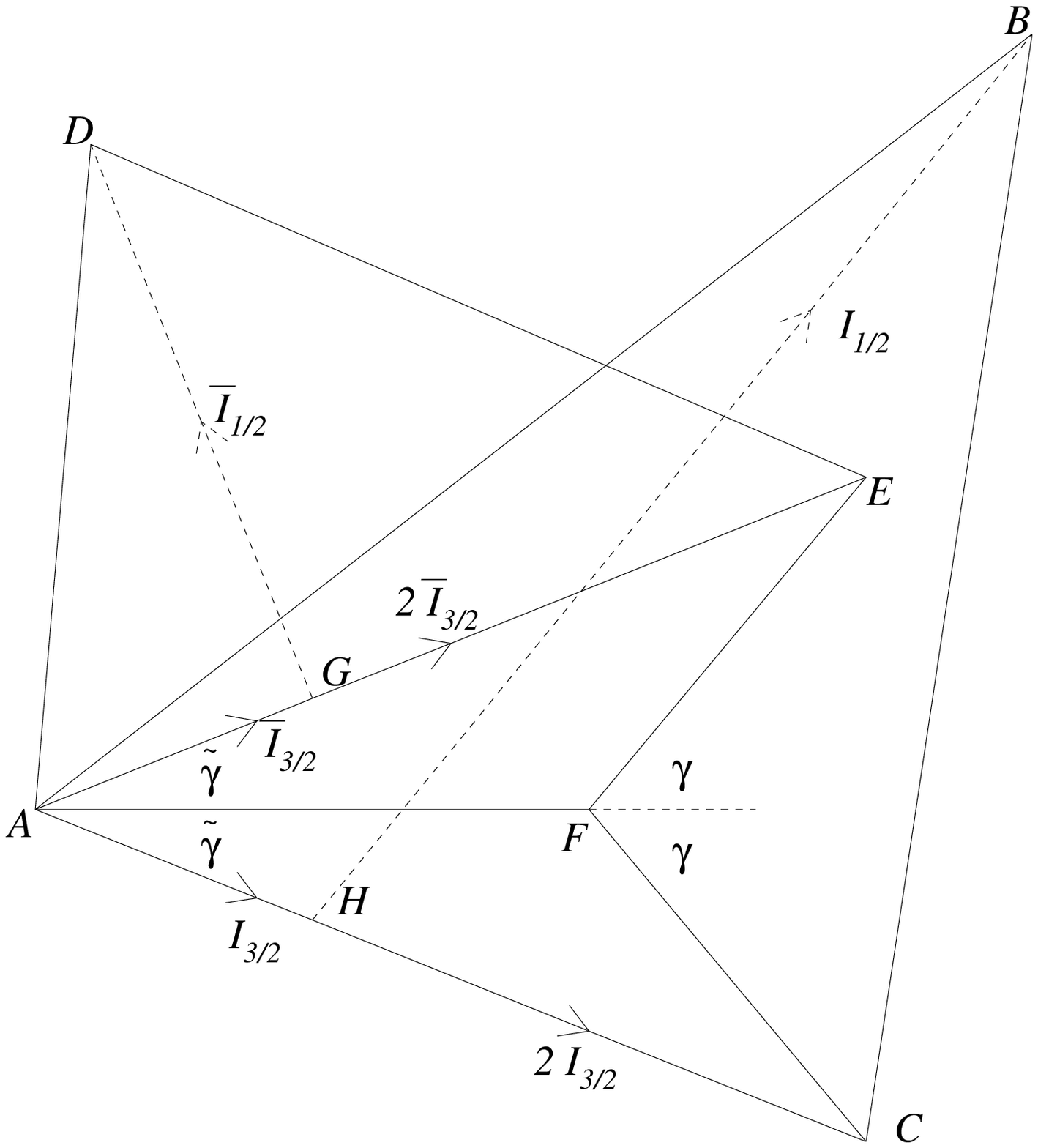}}
\vspace{-0.6in}
\caption{
The triangles formed by the $B_d \rightarrow \pi K$
amplitudes: $AB = | {\cal A} \left( B_d \rightarrow K^+ \pi^- \right)|$,
$BC = |
\protect\sqrt{2} {\cal A}
\left( B_d \rightarrow \pi^0 K^0 \right) |$, $AD =
|{\cal A} \left( \bar{B}_d \rightarrow K^- \pi^+ \right) |
$, $DE = |
\protect\sqrt{2} {\cal A}
\left( \bar{B}_d
\rightarrow \pi^0 \bar{K}^0 \right)| $.
$AF = |
\protect\sqrt{2} \; {\cal A}
\left( B^+ \rightarrow \pi^+ \pi^0 \right)|
\; |\lambda _u ^{(s)}| / |\lambda _u ^{(d)}|$ and $FC = FE =
AF \; \delta_{EW}$ (see Eqn.(\protect\ref{A3/2})).
%Given $\gamma$ and the decay amplitude
%$|B^+ \rightarrow \pi^+ \pi^0|$, $|I_{3/2}|$, $|\bar{{I_{3/2}|$
%and $\tilde{\gamma}$ can be determined using Eqns.(\protect\ref{maga3/2})
%and (\protect\ref{anga3/2}).
As in Fig.\protect\ref{figbpipi}, in the
%coordinate
%system
phase convention
where the strong phase of $C_{15}^T$ is zero, the angle
between $AF$ and the real axis
is $\pi + \gamma$.
}
\protect\label{figbpik}
\end{figure}

Given $\gamma$, we can thus 
construct the triangles of Eqn.(\ref{A3/2}) and its CP-conjugate
(see Fig.\ref{figbpik}). \footnote{
As in section \ref{bpipi},
there is a discrete ambiguity in the orientation of the triangles.}
%$ABC$ and $ADE$ can be eliminated by using the expectations
%that $|I_{1/2} - \bar{I}_{1/2}| \sim |\lambda _c ^{(s)}| P^{\prime}$
%is larger than $|I_{3/2}|$ and the penguin contribution
%dominates in the decay $B_d \rightarrow \pi^0 K^0$ so that
%the angle between $BC$ and $DE$ is close to $2 \; \gamma$.
%Given $\gamma$, we can choose the correct orientation.}
As in section \ref{bpipi}, knowing the magnitudes and orientations of 
$I_{1/2}$
and $\bar{I}_{1/2}$ from Fig.\ref{figbpik}, we can determine
$\tilde{T}^{\prime}$, $P^{\prime}$, 
$\phi^{\prime}_{\tilde{T}}$ and $\phi^{\prime}_P$ 
as functions of $\gamma$, using
equations similar to Eqns.(\ref{idiff1}) and (\ref{idiff2}).

This construction
also shows how to include the EWP contributions
to the $B^+ \rightarrow \pi^+ \pi^0$ decay in section \ref{bpipi} as follows.

Once EWP's are included, as for the case of $I_{3/2}$ and
$\bar{I} _{3/2}$, there
is an angle $2 \; \tilde{\gamma} ^{\prime}$
between $I_{2}$ and $\bar{I} _2$
and the magnitude
of $\tilde{C} _{15}^T$
will depend on
$\gamma$ (see Eqn.(\ref{a2})):
\begin{eqnarray}
3 \; |I_2|  =  3 \; |\bar{I} _2|
& = & \left|\sqrt{2} {\cal A} \left( B^+ \rightarrow \pi^+ \pi^0
\right) \right| \nonumber \\
  & = & 8 |\tilde{C} _{15}^T| |\lambda _u ^{(d)}|
\sqrt{ \left( 1 + \delta ^{\prime \; 2} _{EW} + 2 \delta ^{\prime} _{EW}
\cos \gamma \right)},
\end{eqnarray}
\begin{equation}
\tan \tilde{\gamma}^{\prime} = \frac{\delta ^{\prime} _{EW}
\sin \gamma}{1 + \delta ^{\prime} _{EW} \cos
\gamma},
\end{equation}
where $\delta ^{\prime} _{EW}$ is given by $-|\lambda _c ^{(d)}|
/ |\lambda _u ^{(d)}| \; 3/2 \; \kappa \; / \;
(1-3/2 \; \kappa) \sim O(\hbox{few} \; \%) $.

\subsection{Case 3}
The
$B_d \rightarrow K^+ K^-$
amplitude is given by \cite{d}:
\begin{eqnarray}
{\cal A} \left( B_d \rightarrow K^+ K^- \right) & = & 
- \lambda _u ^{(d)} \left( 2 A^T_3 + 2 A^T_{15} \right)
- \sum _q \lambda _q ^{(d)} \left( 2 A_{3,q}^P + 2 A^P_{15,q} \right) 
\nonumber \\
 & \equiv & a e^{i \phi _a}.
\label{bkk}
\end{eqnarray}
From Eqns.(\ref{b0}), (\ref{b+-}), (\ref{bk+pi-}), (\ref{bk0pi0}) 
and (\ref{bkk}), we can see that 
$\sqrt{2} {\cal A} \left( B_d \rightarrow \pi^0 \pi^0
\right) + {\cal A} \left( B_d \rightarrow
K^- K^+ \right)$ can be obtained from 
$\sqrt{2} {\cal A} \left( B_d \rightarrow K^0 \pi^0
\right)$ and \\
${\cal A} \left( B_d \rightarrow \pi^+ \pi^-
\right) - {\cal A} \left( B_d \rightarrow
K^- K^+ \right)$ can be obtained from ${\cal A} \left( B_d 
\rightarrow \pi^- K^+
\right)$ by scaling the $\Delta S =1$ amplitudes by appropriate CKM factors,
{\it i.e.,} in this case, the decay mode $B_d \rightarrow
K^- K^+$ plays the role of the decay mode $B_s \rightarrow
\pi \pi$ of case \ref{intro} (compare Eqns.(\ref{bspipi}) and (\ref{bkk})).
Thus, as in case \ref{intro}, we
can determine $\gamma$, including {\em all} rescattering effects,
by measuring the  $8$ decay modes: $B^+ \rightarrow \pi^+ \pi^0$,
$B_d$ and $\bar{B}_d
\rightarrow \pi K$ (all), $B_d \rightarrow K^+ K^-$,
$B_d \rightarrow \pi^0 \pi^0$ and $B_d \rightarrow \pi^- \pi^+$  (or
CP-conjugates of the last three modes). 
%As mentioned 
%in case \ref{intro}, if the
%annihilation amplitudes are small, then 
%the decay amplitudes
%for $B_d \rightarrow \pi K$ and $B_d \rightarrow \pi \pi$ are
%the same, up to CKM factors,
%which can be used in this case to predict a $B_d \rightarrow \pi \pi$
%decay rate as a function of $\gamma$. Thus, we can
%determine $\gamma$ by measuring
%any {\em one} $B_d \rightarrow \pi \pi$ decay mode, in addition
%to the $B^+ \rightarrow \pi^+ \pi^0$, $B_d$ (and $\bar{B}_d$)
%$\rightarrow \pi K$ decay 
%modes.  
If the
annihilation amplitudes are small, as in case \ref{intro}, we can
determine $\gamma$ by measuring
any {\em one} $B_d \rightarrow \pi \pi$ decay mode, in addition
to the $B^+ \rightarrow \pi^+ \pi^0$, $B_d$ (and $\bar{B}_d$)
$\rightarrow \pi K$ decay
modes. As in case \ref{intro}, if we measure the CP-conjugate
$B_d \rightarrow \pi \pi$ rates as well, then a CP-averaged measurement
of the decay rate $B_d \rightarrow K^+ K^-$ suffices.

\subsection{Case 4}
\label{bspik}
The expressions for the decay amplitudes for 
$B_s \rightarrow \pi^+ K^-$ and $B_s \rightarrow \pi^0 \bar{K}^0$
in terms of the $SU(3)$ invariant amplitudes
are identical to those
for $B_d \rightarrow \pi^- K^+$ and $B_d \rightarrow \pi^0 K^0$, 
respectively,
{\em including} annihilation contributions
(unlike the case of $B_d \rightarrow \pi \pi$ and
$B_d \rightarrow \pi K$ decays), modulo the CKM factors \cite{d}.
Thus, the same method predicts the rates for the $B_s \rightarrow 
\pi K$ decays and the CP-conjugate processes.
It suffices to use the $B_s \rightarrow K^- \pi^+$
decay (or its CP-conjugate) which is a ``self tagging'' mode
or the CP-averaged decay rate for $B_s \rightarrow \pi^0 \bar{K}^0$. Thus, no
external tagging is required.
%As mentioned in the introduction, Gronau and Pirjol \cite{gp}
%also gave a method to determine $\gamma$ using 
%the
%$B_s \rightarrow \pi K$ decays (using the fact that
%these decay amplitudes are similar to the $B_d \rightarrow 
%\pi K$ amplitudes), but they require measurements of
%{\em all} the
%$B_s \rightarrow \pi K$ decays.

\section{Discussions}
\label{conclude}
The analysis of the above three cases is strictly valid
only in the flavor $SU(3)$ limit. In the tree level
amplitudes, {\it i.e.,}
$C_{15}^T$ and the $\tilde{T}$'s, the 
corrections due to 
$SU(3)$ breaking can be taken into account 
more reliably in the factorization approximation
and are expected to be given 
by $f_K/f_{\pi}$ (see, for example, Gronau {\it et al.} in
\cite{ghlr1} \cite{nq}). For example, in Fig.\ref{figbpik} the lengths
of $AF$, $FE$ and $FC$ will have to be multiplied by $f_K/f_{\pi}$.
However, since the strong penguin amplitudes include $(V-A) (V+A)$
type operators, the (factorizable)
corrections due to $SU(3)$ breaking there
are less certain, but the corrections are still less than
$\sim O(30\%)$. This is especially relevant for the cases 
\ref{intro} and \ref{bpipi}
where the penguin contribution dominates
in the $B_d \rightarrow \pi K$, $B_s \rightarrow
K^+ K^-$ decays and we are {\em predicting}
this contribution from the $B_d \rightarrow \pi \pi$ decays
using $SU(3)$ symmetry.
In the cases \ref{bpik} and \ref{conclude}, we use $SU(3)$ symmetry to 
predict the penguin contribution in a $B_d \rightarrow \pi 
\pi$ (or a $B_s \rightarrow \pi K$) decay
from the $B_d \rightarrow \pi K$ decays, but now the {\em tree level} 
contributions
dominate the $B_d \rightarrow \pi \pi$ decay rate and so the uncertainty
due to the $SU(3)$ breaking in the penguin amplitudes is less
important.

We have a prediction for more than one rate in some of the
cases. For example, in case \ref{conclude}, we can predict 
both $B_s \rightarrow \pi^+ K^-$ and its CP-conjugate decay rate
or in case \ref{intro}, neglecting annihilation, we can predict
the decay rates $B_d \rightarrow \pi^- K^+$ and $\pi^0 K^0$.
So, we can treat the $SU(3)$ breaking in the penguin amplitudes as an
unknown and determine it (in addition to $\gamma$) from
the measurement of {\em seven} decay rates.

We have also assumed that the $SU(3)$ breaking in the strong phases
is small. A possible justification is that at the energies
of the final state particles $\sim m_b/2$, the phase shifts are
not expected to be sensitive to the $SU(3)$ breaking given by,
say, $m_K - m_{\pi}$ (which is much smaller than
the final state momenta). However, it is hard to
quantify this effect.
 
If we measure {\em all} the $B \rightarrow \pi \pi$
and $B_d \rightarrow \pi K$ decay rates, then we can compute
the tree level
parts of the amplitudes, both $\tilde{T}$ and $\tilde{T}^{\prime}$ 
(see Eqns.(\ref{a0}) and (\ref{a1/2})),
as functions
of $\gamma$ as discussed in sections \ref{bpipi} and \ref{bpik}. 
If the annihilation amplitudes are small,
then we have $\tilde{T} = \tilde{T}^{\prime}$ (in the $SU(3)$ limit) since
the decay amplitudes
$B_d \rightarrow \pi \pi$ and $B_d \rightarrow \pi K$
are the same up to CKM factors. 
To  
include the
$SU(3)$ breaking in this analysis, we 
use the modified relations
$C_{15}^T \; (\Delta S =1) = f_K/f_{\pi} \;
C_{15}^T \; (\Delta S =0)$ 
and $\tilde{T}^{\prime}
= f_K/f_{\pi} \; \tilde{T}$. This can be used to determine
$\gamma$, including $SU(3)$ breaking (without having to 
to deal with $SU(3)$ breaking in the penguin amplitudes
and in the strong phases).

\end{document}